\documentclass[12pt,preprint]{emulateapj}

\usepackage{graphicx}
\usepackage{amsmath}
\usepackage{hyperref}
\usepackage{color}

\hypersetup{
    colorlinks=true,
    linkcolor=red,
    citecolor=blue,
    urlcolor=magenta,
}

\begin{document}

\title{Impact of Microlensing on Observational Strategies for Gravitational Time Delay Measurements}
\author{Kai Liao$^{1}$}

\affil{
$^1$ {School of Science, Wuhan University of Technology, Wuhan 430070, China}\\
}

\begin{abstract}
  Microlensing not only brings extra magnification lightcurves on top of the intrinsic ones but also shifts them in time domain,
  making the actual time-delays between images of strongly lensed active galactic nucleus change on
  the $\sim$ day(s) light-crossing time scale of the emission region. The microlensing-induced time-delays would bias strong lens time-delay cosmography
  if uncounted. However, due to the uncertainties of the disk size and the disk model, the impact is hard to accurately estimate.
  In this work, we study how to reduce the bias with designed observation strategy based on a standard disk model. We find long time monitoring of the images could
  alleviate the impact since it averages the microlensing time-lag maps due to the peculia motion of the source relative to the lens galaxy.
  In addition, images in bluer bands correspond to smaller disk sizes and therefore benefit time-delay measurements as well.
  We conduct a simulation based on a PG 1115+080-like lensed quasar.
  The results show the time-delay dispersions caused by microlensing can be reduced by $\sim40\%$ with 20-year lightcurves
  while u band relative to r band reduces $\sim75\%$ of the dispersions.
  Nevertheless, such an effect can not be totally eliminated in any cases.
  Further studies are still needed to appropriately incorporate it in inferring an accurate Hubble constant.

\end{abstract}

\section{Introduction}
The issue on the Hubble constant ($H_0$) is one of the severest concerns for current cosmology since the value inferred from
the early-time cosmic microwave background (CMB) observations~\citep{Planck2018} is inconsistent with that measured by the late-time distance ladders~\citep{Riess2019}.
The discrepancy implies either unknown systematic errors or new physics. A third-party approach is expected to bring new perspectives and be the referee.

Strong lensing by galaxies is one of such approaches~\citep{Treu2016}. The typical system consists a distant quasar lensed by the foreground elliptical galaxy, forming multiple images
of the active galactic nucleus (AGN) arriving at the earth in turn. The time-delay between any two of the images is inversely proportional to $H_0$ and weakly
depends on other cosmological parameters as well. With the lens potential information from high-resolution imaging of the host galaxy and
spectroscopics, time-delay measurements can be used to infer $H_0$.
The state-of-art lensing program H0LiCOW has achieved a $2.4\%$ precision of $H_0$ measurement with time-delay method~\citep{Wong2020}
in agreement of the results from distance ladders. Inspired by the achievement, the new collaboration team TDCOSMO~\citep{Millon2020} will go further to measure
$H_0$ with $1\%$-level precision. However, before combining more lenses to get more precise constraint,
factors that impact the accuracy of the inference should be addressed, i.e., accuracy is more important than pursing precision.

Inferring $H_0$ with time-delay lensing needs at least three ingredients: 1) lens potentials; 2) time-delays; 3) mass density fluctuations along the line-of-sight.
Challenges on the accuracy of each one were proposed. For example, the lens modelling may bring dominated bias~\citep{Schneider2013,Birrer2016,Kochanek2020,Ding2020}.
For the time-delay measurements, the Time Delay Challenge (TDC) program~\citep{Liao2015} proved the bias can be well-controlled, i.e., measurements
with lightcurves contaminated by the magnification patterns of microlensing are accurate.
However, such a time-delay measurement was recently pointed out not to be
the cosmologically concerned one~\citep{TK18} and it depends on the observing band~\citep{Liao2020}.

Time-delays are measured by comparing the phases of the lightcurve pairs.
In fact, gravitational microlensing also produces changes in the actual time-delays on the $\sim$day(s) light-crossing
time-scale of the emission region of the accretion disk. This effect is due to a combination of the inclination of the disk
relative to the line of sight and the differential magnification of the temperature fluctuations producing the variability~\citep{TK18} which changes the
arriving timings of the lightcurve phases.
Directly verifying such an effect can be done by measuring time-delay ratio anomalies or
time-delay difference between bands~\citep{Liao2020}.
With multi-band light curves, lensed quasars displaying multiple AGN images
can be used to effectively measure disk sizes, though each image is affected by
its own microlensing map~\citep{Liao2020,Chan2020}. Note that while ~\citep{Liao2020} focused on time-delays between images,
~\citep{Chan2020} studied time-lags for each image via ``micro-lensed" reverberation mapping. To avoid confusion, we use the term ``time-lag" for single image while ``time-delay" for that between lensed images
throughout this paper.

For cosmological studies, this effect would bias $H_0$ inference if uncounted, especially for small time-delays.
In principle, one can get the microlensing time-delay distributions and incorporate them within the Bayesian framework~\citep{Chen2018} (though the time-delays at different epoches should be correlated
in their analysis).
However, the most uncertain factor comes from the
disk size (and disk model itself) which has been found to be larger than that predicted by the standard thin-disk model
of AGN with reverberation mapping of the continuum emission and microlensing observations~\citep{Collier1998,Sergeev2005,Shappee2014,Fausnaugh2016,Jiang2017,Pooley2007,Mosquera2013,MacLeod2015,Li2019,Cornachione2020}.
It is therefore impossible to correctly understand the prior distributions for current techniques.
The existing analysis has not considered such an effect due to not finding time-delay variations with observing time (but the effect may still exist)~\citep{Bonvin2018,Wong2020}.
Another reason is the H0LiCOW team has chosen the largest time-delays such that the relative errors become negligible. However, systems like lens PG 1115+080
which has short time-delays would be affected. Ignoring such an effect would limit the lens selection.
Note that most works in the literature considered a motionless source which is not realistic~\citep{Chen2018,Bonvin2018}.
The source would travel along a line in the microlensing time-lag maps with finite monitoring time
due to its peculia motion relative to the lens galaxy.
The microlensing time-delay effect is therefore supposed to be averaged.

In this work, rather than discussing how to correctly estimate the impact by microlensing time-delays,
we study the possibility that increasing the monitoring time in bluer bands can reduce the impact such that the $H_0$ inference can be accurate to the greatest extent without considering such an effect.
The paper is organized as follows: In Section 2, we introduce the chromatic microlensing time-delay effect; In Section 3, we discuss the finite lightcurve case;
Simulations and results are presented in Section 4; Finally, we summarize and make discussions in Section 5.

\section{Chromatic microlensing time delays}
Though details of accretion disk models are being debated, a simple thin-disk model is widely concerned as the standard one~\citep{Shakura1973}.
For a non-relativistic, thin-disk model that emits as a blackbody, the characteristic size of the disk is defined as
\begin{equation}\label{R0}
\begin{aligned}
R_0&=\left[\frac{45G\lambda_\mathrm{rest}^4M_\mathrm{BH}\dot{M}}{16\pi^6h_\mathrm{p}c^2}\right]^{1/3} \\
&=9.7\times10^{15}\left(\frac{\lambda_\mathrm{rest}}{\mathrm{\mu m}}\right)^{4/3}\left(\frac{M_\mathrm{BH}}{10^9M_\odot}\right)^{2/3}\left(\frac{L}{\eta L_\mathrm{E}}\right)^{1/3}\mathrm{cm},
\end{aligned}
\end{equation}
which corresponds to the radius where the temperature matches the photon rest-frame wavelength, i.e., $kT=h_\mathrm{p}c/\lambda_\mathrm{rest}$.
$k,G,h_\mathrm{p},c$ are Boltzmann, Newonian, Planck constants and speed of light, respectively. $M_\mathrm{BH}$ is the mass of the black hole.
$L/L_\mathrm{E}$ is the fractional luminosity with respect to the Eddington luminosity.
$\eta=L/\dot{M}c^2$ is the accretion efficiency ranging from $\sim0.1$ to $\sim0.4$ that positively correlates with the black hole spin.
Note that adjusting to $\eta$ can not result in the observed larger disk size with current measurement precision.
The corresponding characteristic time scale:
\begin{equation}
\frac{(1+z_s)R_0}{c}\simeq \frac{3.8\ \mathrm{days}}{(1+z_s)^{1/3}}\left(\frac{\lambda_\mathrm{obs}}{\mathrm{\mu m}}\right)^{4/3}\left(\frac{M_\mathrm{BH}}{10^9M_\odot}\right)^{2/3}\left(\frac{L}{\eta L_\mathrm{E}}\right)^{1/3}.
\end{equation}

It is convenient to define a dimensionless radius:
\begin{equation}\label{xi}
\xi=\frac{h_\mathrm{p}c}{kT_0(R)\lambda}=\left(\frac{R}{R_0}\right)^{3/4}\left(1-\sqrt{\frac{R_\mathrm{in}}{R}}\right)^{-1/4},
\end{equation}
where $R_\mathrm{in}$ is the inner edge of the disk~\citep{Morgan2010}, depending on whether it is a Schwarzschild or Kerr black hole.
The unperturbed temperature profile and surface brightness profile of the disk are given by $T_0(R)^4\propto R^{-3}(1-\sqrt{R_\mathrm{in}/R})$
and $I_0(R)\propto (e^\xi-1)^{-1}$, respectively.

For the variability, in a ``lamp post" model~\citep{Cackett2007}, the fractional temperature variation is independent of the position in the disk.
The disk center acts as the driving source, resulting in variability lagged by the light travel time $R/c$. Assuming the variations
are small, the blackbody function can be Taylor expanded and the brightness variability is given by
\begin{equation}
\delta I(R,t)\propto f(t-R/c)G(\xi),
\end{equation}
where
\begin{equation}\label{G}
G(\xi)=\frac{\xi e^\xi}{(e^\xi-1)^2}.
\end{equation}

Note that what one can measure is the phase difference of the light curves between lensed images as the time-delay $\Delta t_\mathrm{lc}$.
Microlensing makes different parts of the disk contribute differently, inducing different time-lags for each image (we refer ~\citep{TK18} for more details).
The mean time-lag caused by microlensing is given by~\citep{TK18,Liao2020}
\begin{equation}
t_\mathrm{micro} = \frac{1+z_s}{c}\frac{\int dudvG(\xi)M(u,v)R(1-\cos\theta \sin \beta)}{\int dudvG(\xi)M(u,v)}-t_\mathrm{disk},\label{tmicro}
\end{equation}
where $M(u,v)$ is the microlensing magnification map projected in the source plane, $\theta$ is the polar angle in the accretion disk plane,
$\beta$ is the inclination angle with $\beta=0$ corresponding
to a face-on disk. $u=R\cos\theta\cos \beta$ and $v=R\sin\theta$ are the coordinates in the source plane.
The \emph{geometric delay} appearing in the reverberation mapping method by the disk itself is
\begin{equation}\label{diskt}
t_\mathrm{disk}=\frac{1+z_s}{c}\frac{\int dudvG(\xi)R(1-\cos\theta \sin \beta)}{\int dudvG(\xi)},
\end{equation}
which corresponds to the case without microlensing. It cancels out between lensed images.
Ignoring the inner edge of the disk, $t_\mathrm{disk}=5.04(1+z_s)R_0/c$. In Eq.\ref{tmicro} the integral radius is from $R_\mathrm{in}$ to infinity. In practice,
a $30R_0$ region is sufficient. We set $R_\mathrm{in}=R_0/100$ whose specific value impact little. The microlensing time-lag primarily depends on
the disk size, and weakly depends on the view angles as well~\citep{TK18,Bonvin2018,Liao2020}.
The measured time-delay relative to the cosmological one is therefore changed by $\Delta t_\mathrm{lc}=\Delta t_\mathrm{cosm}+\Delta t_\mathrm{micro}$.

Moreover, the microlensing time-delays was proposed to be chromatic~\citep{Liao2020} since the disk size is a function of wavelength, i.e., time-delays measured with lightcurves in different bands
are different: $\delta\Delta t_\mathrm{lc}\neq0$. Bluer bands would have smaller effects. Since multiple concepts of time exist, to avoid confusion,
we have adopted the sign convention made by~\citep{Liao2020}: $t$ without prefix denotes time-lag relative to the driving source $f(t)$ for single image, $\Delta$ is
for difference between images (time-delay), $\delta$ is for difference between bands and $\mathcal{T}$ denotes the observing time.

\begin{figure*}
\centering
\includegraphics[width=5cm,angle=0]{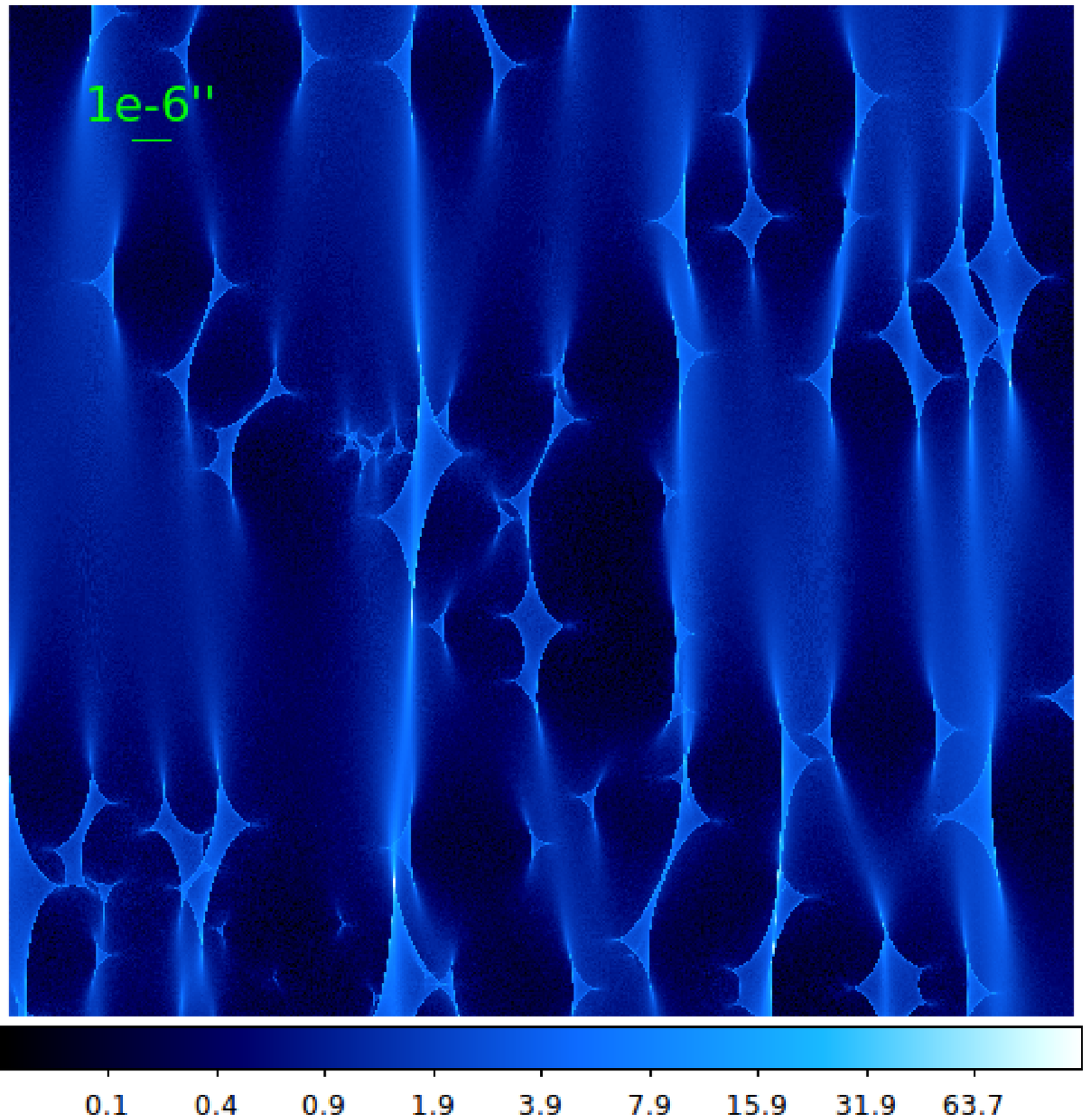}
\includegraphics[width=5cm,angle=0]{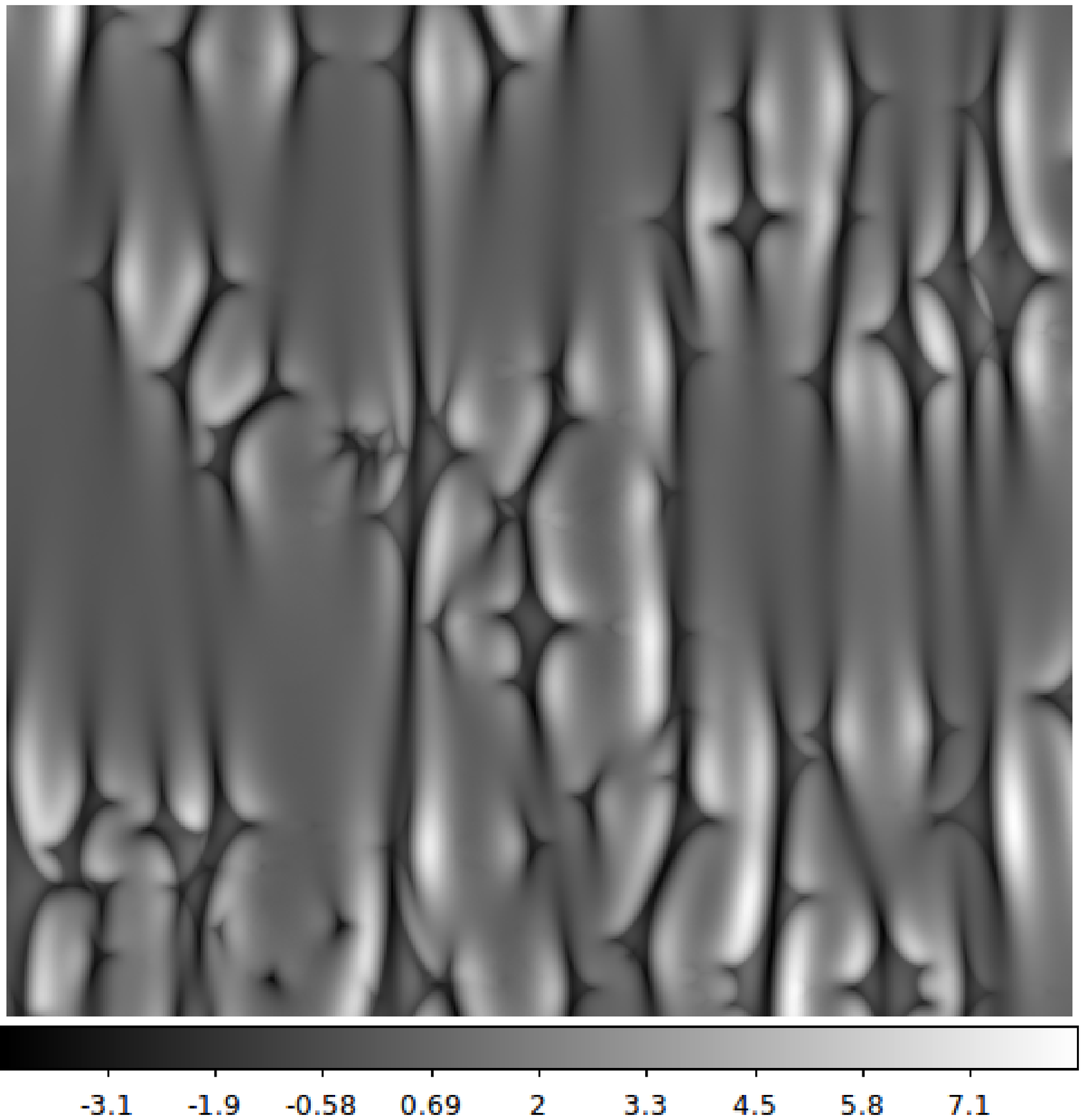}
\includegraphics[width=5cm,angle=0]{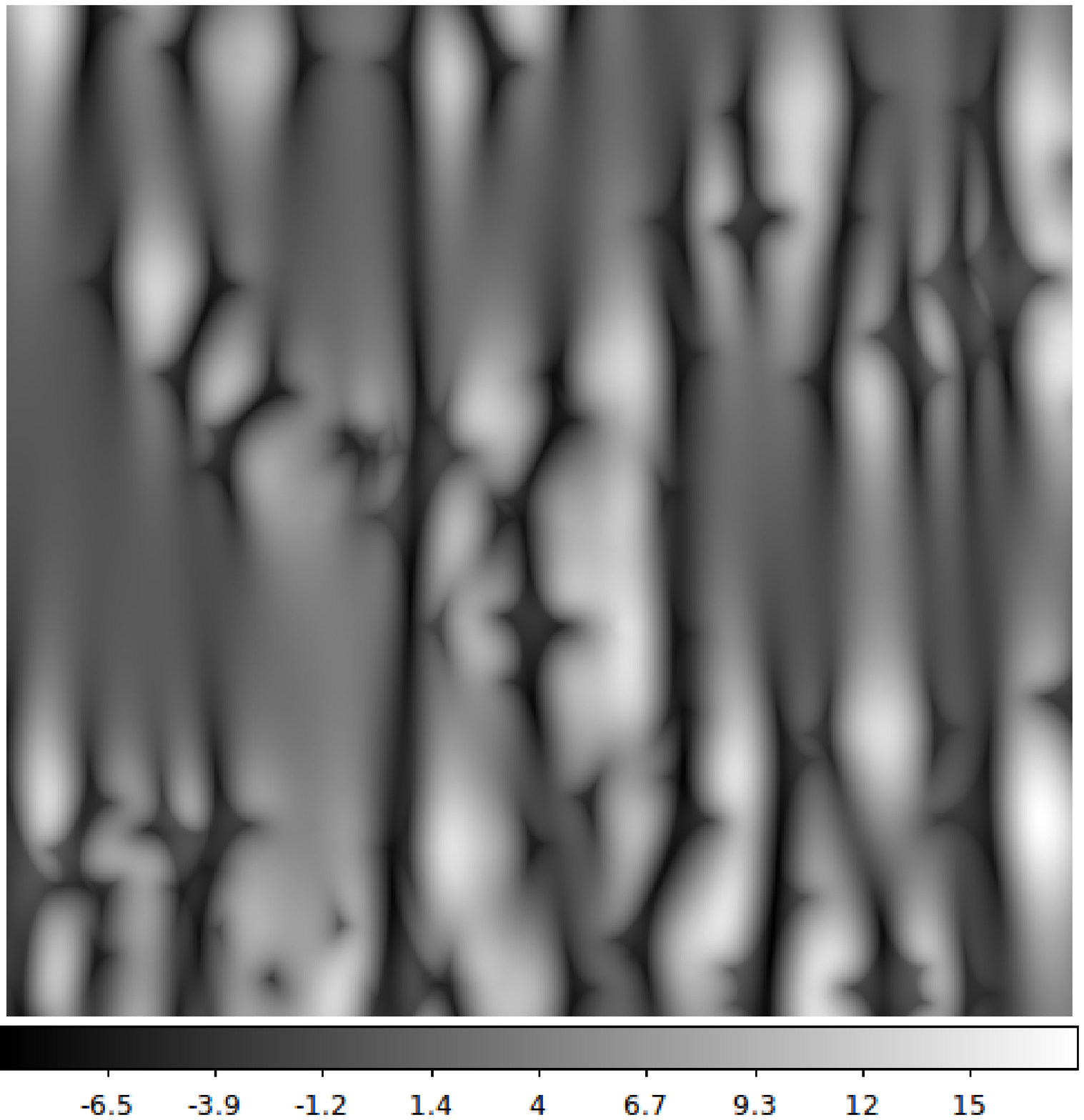}
\includegraphics[width=5cm,angle=0]{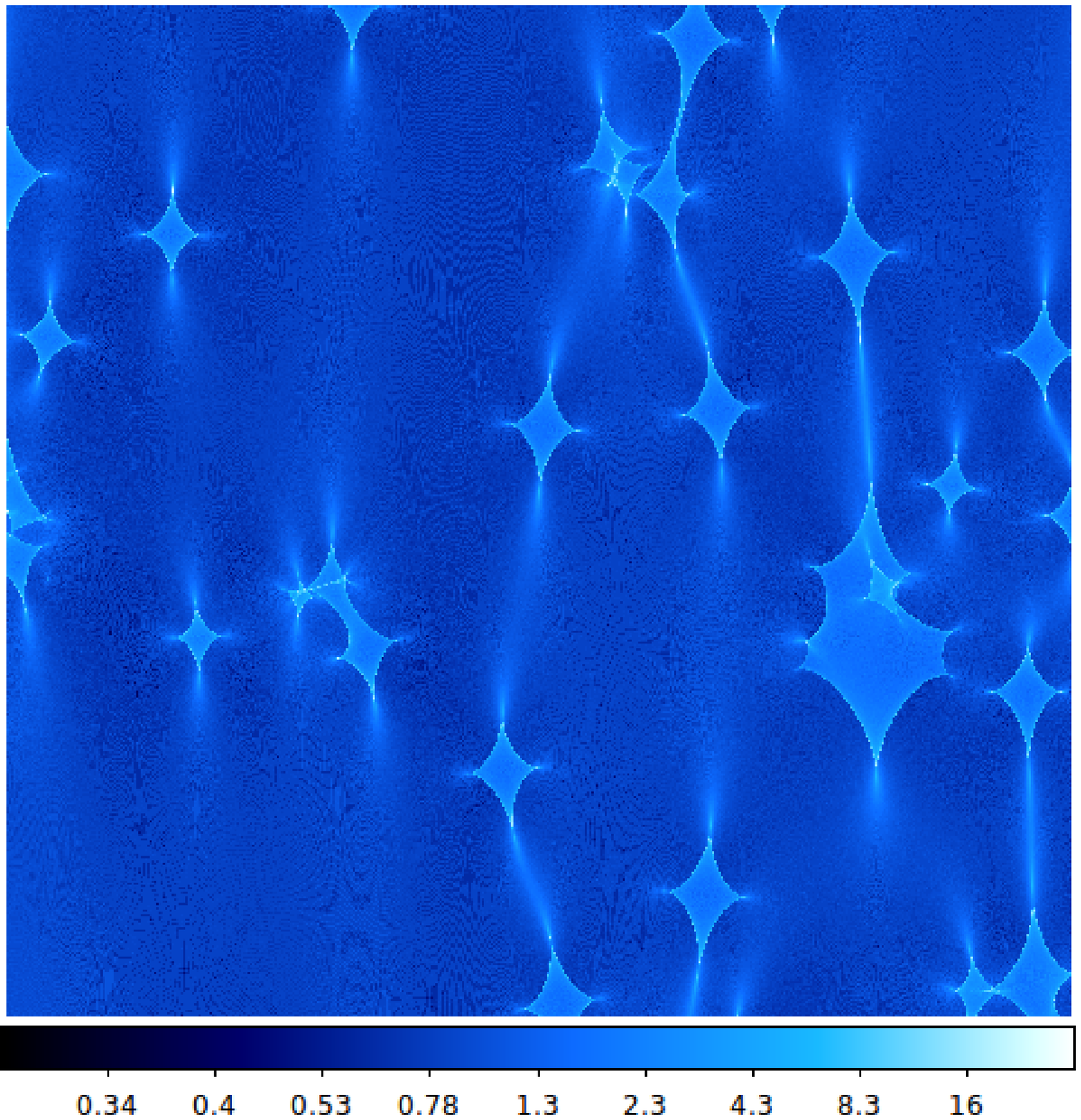}
\includegraphics[width=5cm,angle=0]{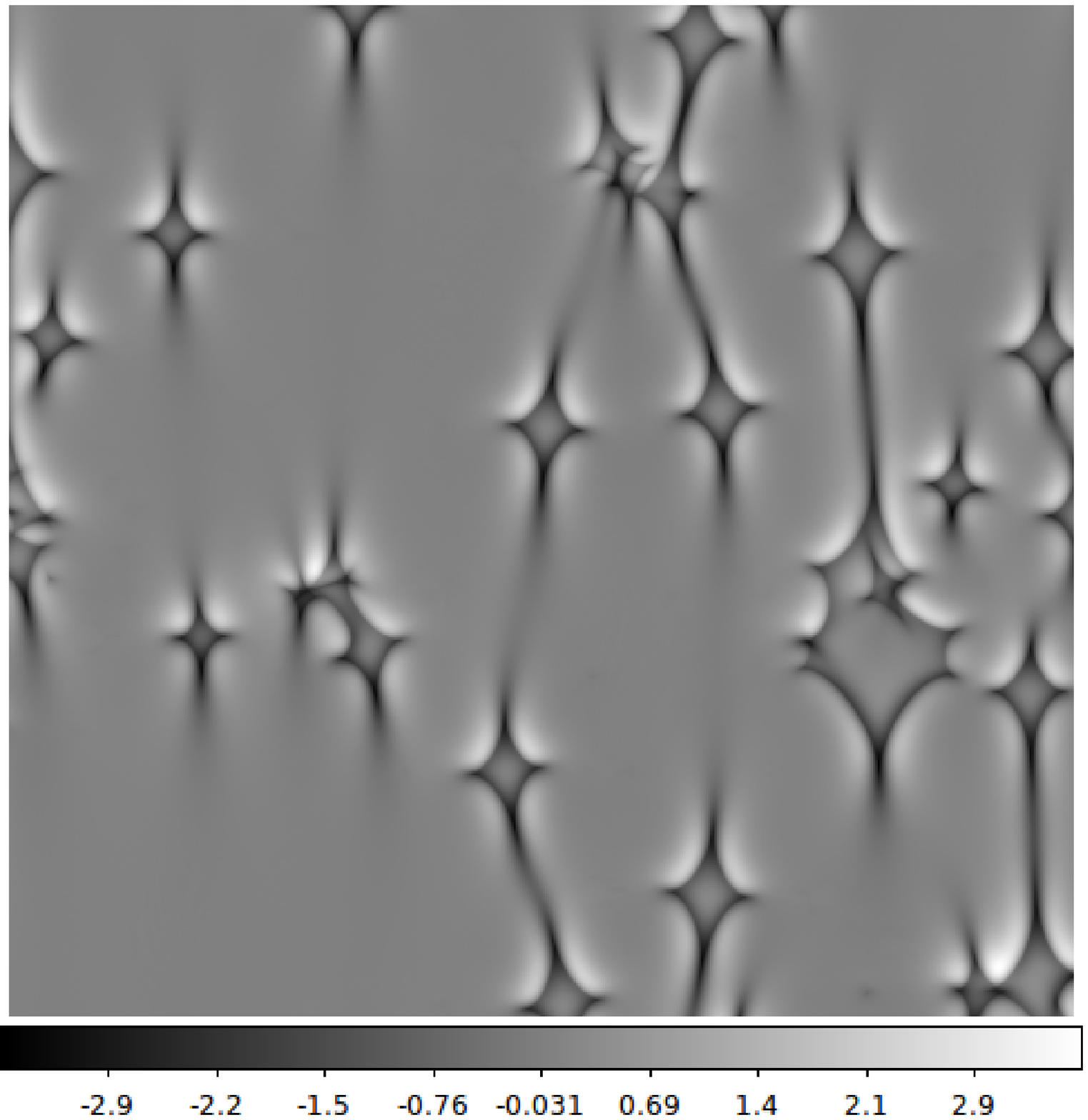}
\includegraphics[width=5cm,angle=0]{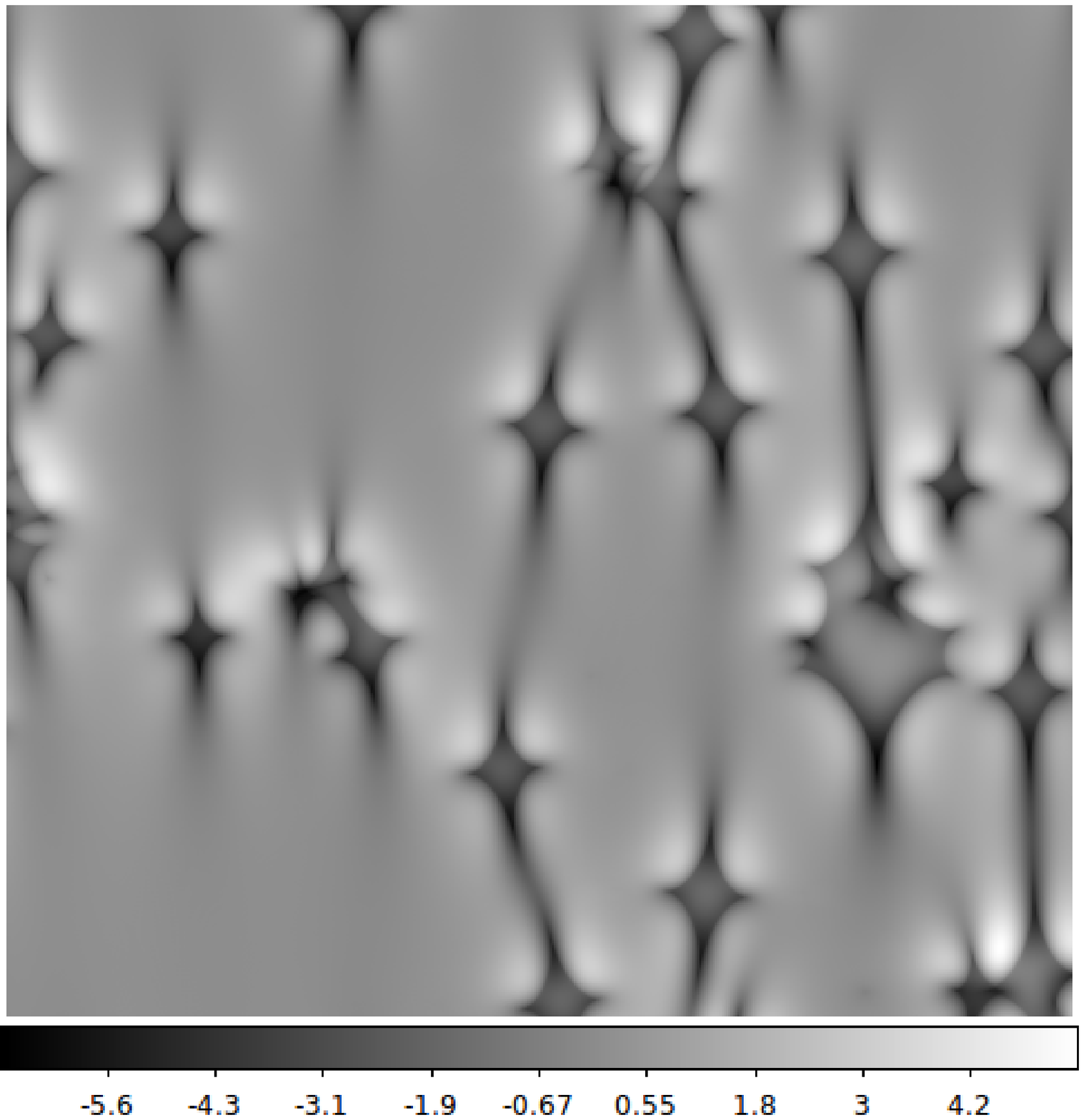}
\caption{Maps for the two images of PG 1115+080 (one of the realizations).
The blue ones are microlensing magnification maps in units of the magnification value while the grey ones are microlensing time-lag maps in units of day. The upper and lower panels are for image B and C, respectively.
 The middle column corresponds to the standard thin-disk model with disk size $R_0$ while the right column is for the non-standard model with disk size $2R_0$. All maps are squares with the same dimension
 showed in the first one.
}\label{map}
\end{figure*}

\section{Finite lightcurve case}
The mean values of the microlensing time-delay lag maps were found not to be vanished~\citep{TK18}, suggesting one can not remove such an effect by monitoring the lens for a
long period of time. Nevertheless, we conjecture that longer lightcurves may reduce the impact to a level that does not bias $H_0$ comparing to uncertainties from other aspects.
The disk will cover more parts of the magnification maps with the peculia motion.
Microlensing time-delays will therefore slowly change~\citep{TK18}. Though the changing is hard to be seen
since the time scales required must be larger than that typically gives a time-delay
measurement $\sim$year(s)~\citep{Liao2020}, it actually implies that finite lightcurves could average such an effect.

Microlensing time-lags actually depend on observed band and observing epoch $t_\mathrm{micro}(b,\mathcal{T})$, so does the measured time-delay $\Delta t_\mathrm{lc}(b,\mathcal{T})$.
The mean value of the measured time-delay is
\begin{equation}
\Delta t^\mathrm{mean}_\mathrm{lc}(b)=\frac{\int\Delta t_\mathrm{lc}(b,\mathcal{T})d\mathcal{T}}{\int d\mathcal{T}},\label{meant}
\end{equation}
which is the one reported by the time-delay programs like the COSMOGRAIL~\footnote{www.cosmograil.org}. Note that various algorithms used to give time-delay measurements generally work in this way~\citep{Liao2015}:
shifting the lightcurves of the lensed images both in flux and time and finding the best-fit value such that the fitness reaches the maximum.
During this process, each piece of the light curves (each segment of a long monitoring campaign) are equally considered, i.e., contributing equally.
Suppose that each piece gives different time-delays,
the overall best-fit value will be approximately the mean time-delay.
In the next section, we will conduct a simulation to prove that the dispersion of $\Delta t^\mathrm{mean}_\mathrm{lc}(b)$ would be smaller than that of $t_\mathrm{lc}(b,\mathcal{T})$.
Besides, bluer bands like u band with shorter wavelengths will give smaller dispersions than redder bands like r band.

\begin{figure*}
\centering
\includegraphics[width=5cm,angle=0]{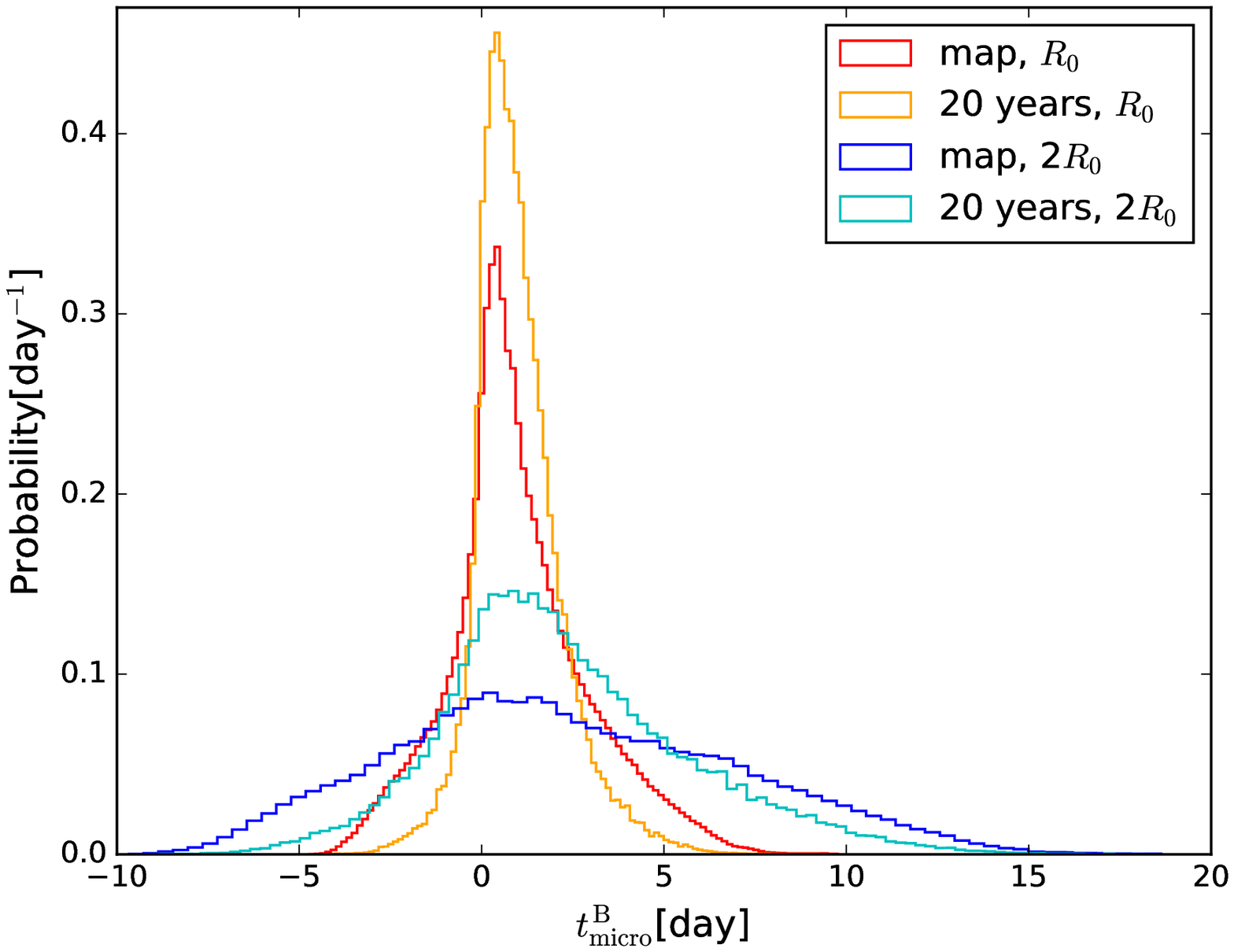}
\includegraphics[width=4.9cm,angle=0]{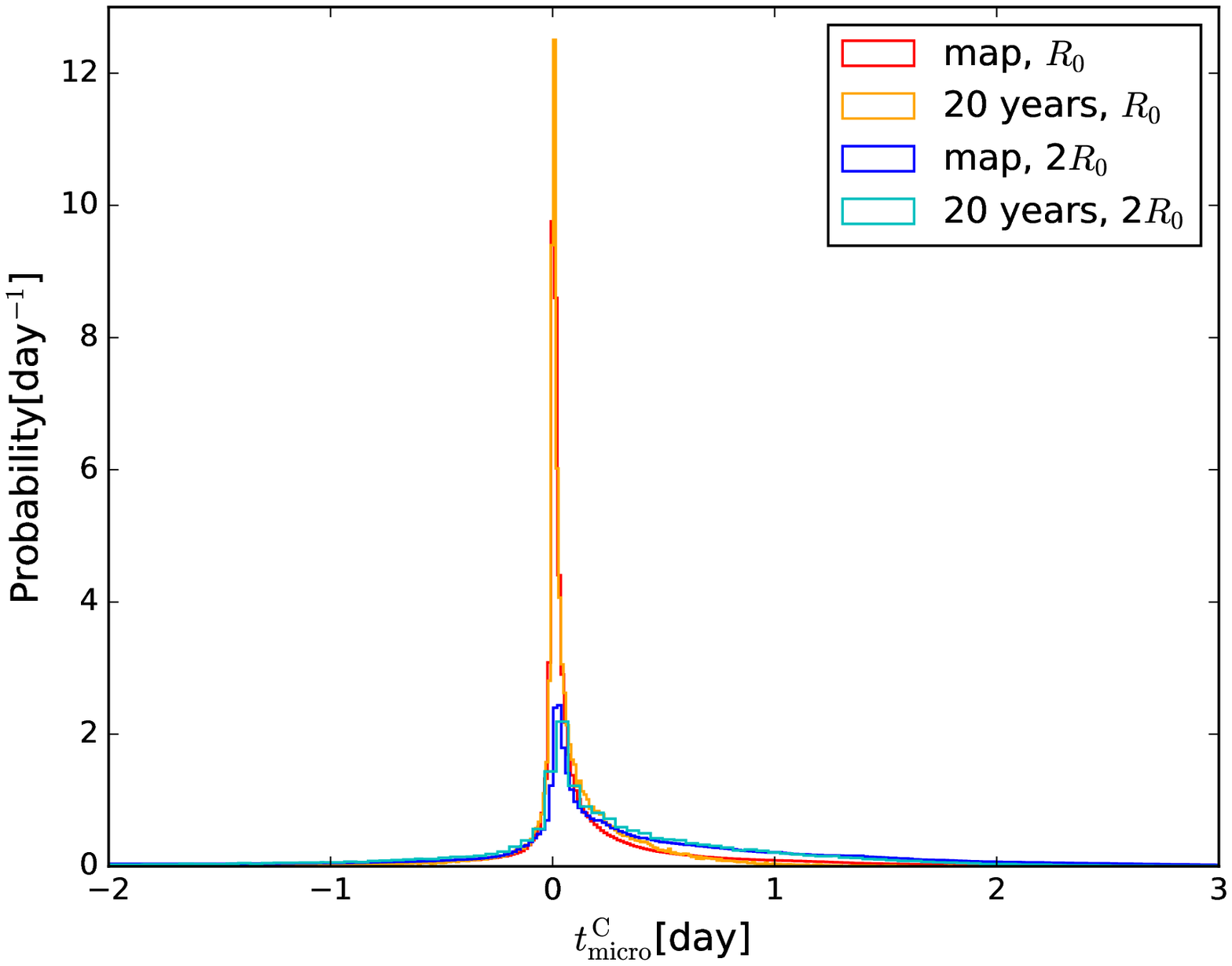}
\includegraphics[width=5cm,angle=0]{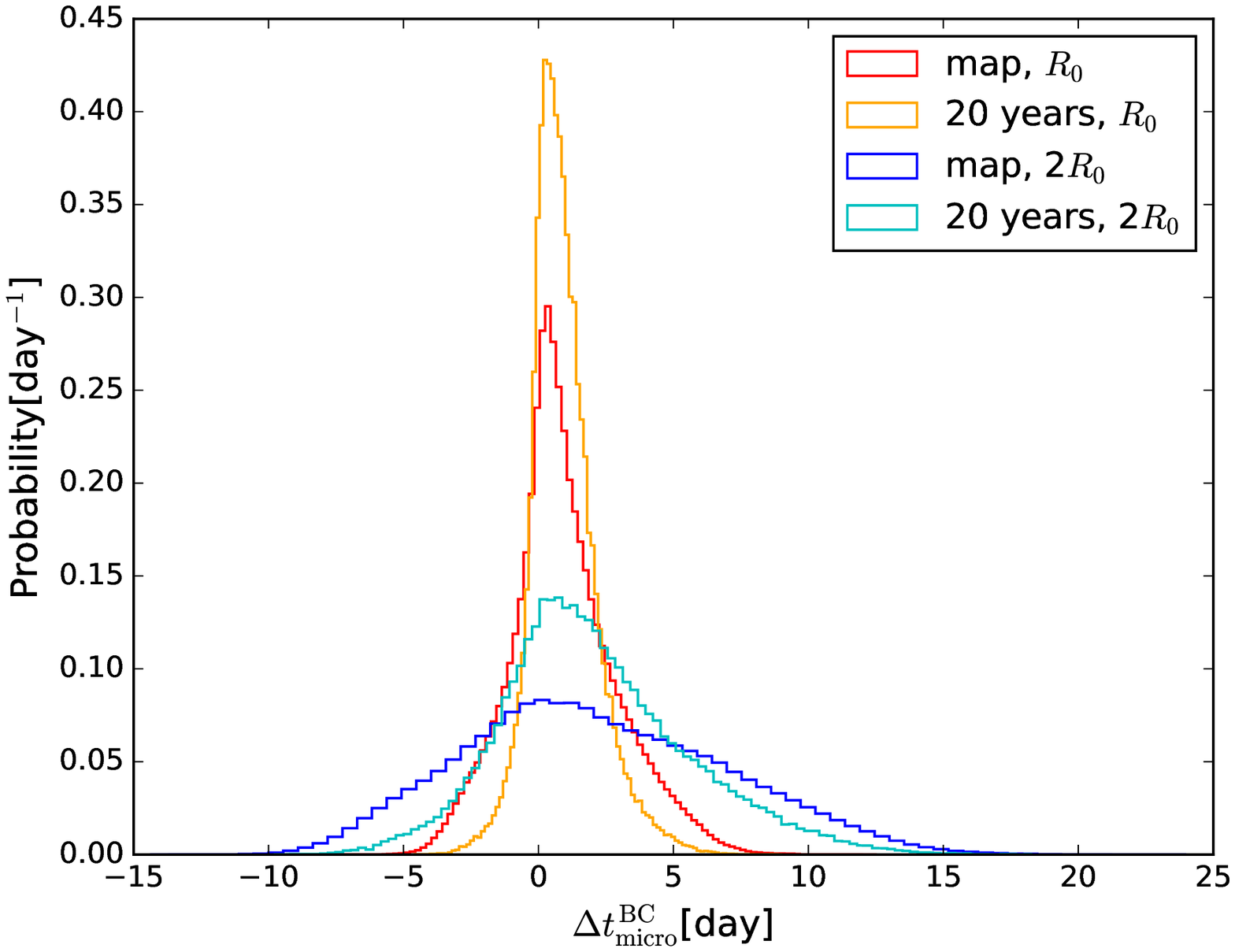}
\caption{Normalized probability distributions of the microlensing time-lags for the two images (left and middle panels) and microlensing time-delay between them (right panel) in r band. Disk sizes $R_0$ and
2$R_0$ are considered, respectively. The red and blue lines termed ``map" correspond to the motionless case while ``20 years" is for finite monitoring.
}\label{tdis}
\end{figure*}

\section{Simulations and results}

\begin{figure*}
\centering
\includegraphics[width=6.85cm,angle=0]{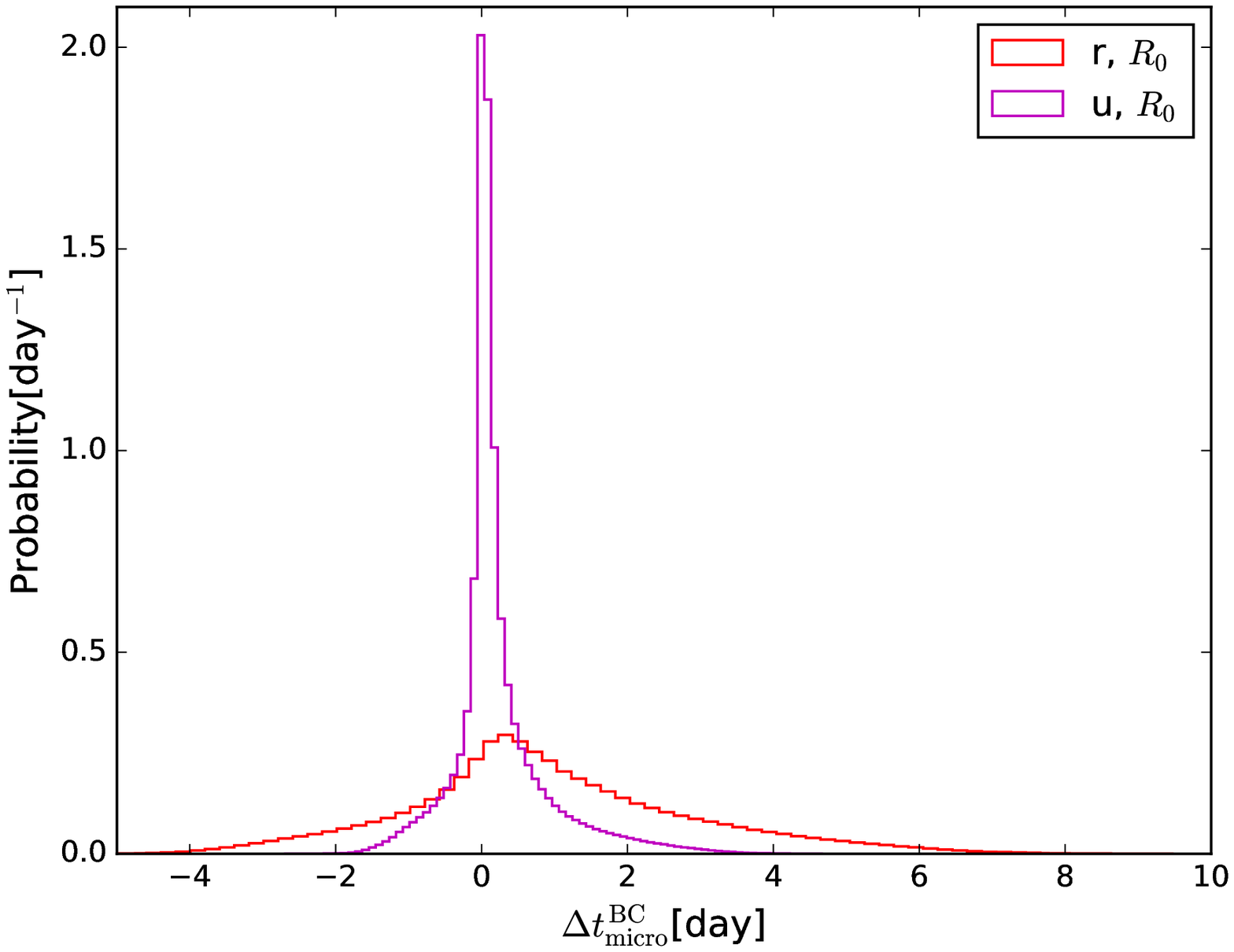}
\includegraphics[width=7cm,angle=0]{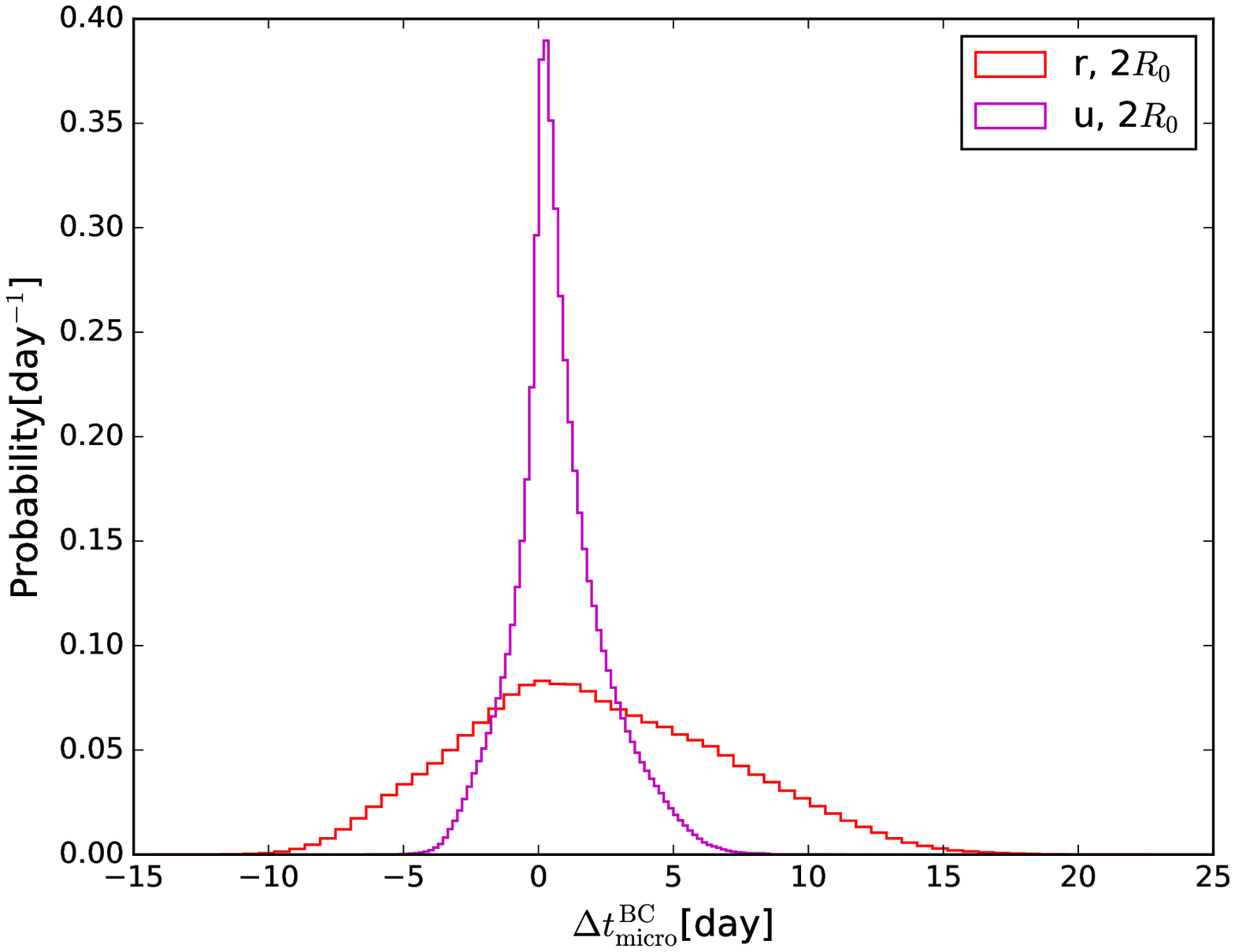}
\caption{Normalized probability distributions of the microlensing time-delays between two images B and C. We show
 both r band and u band for disk sizes $R_0$ and $2R_0$, respectively.}\label{ru}
\end{figure*}

\begin{figure*}
\centering
\includegraphics[width=5.15cm,angle=0]{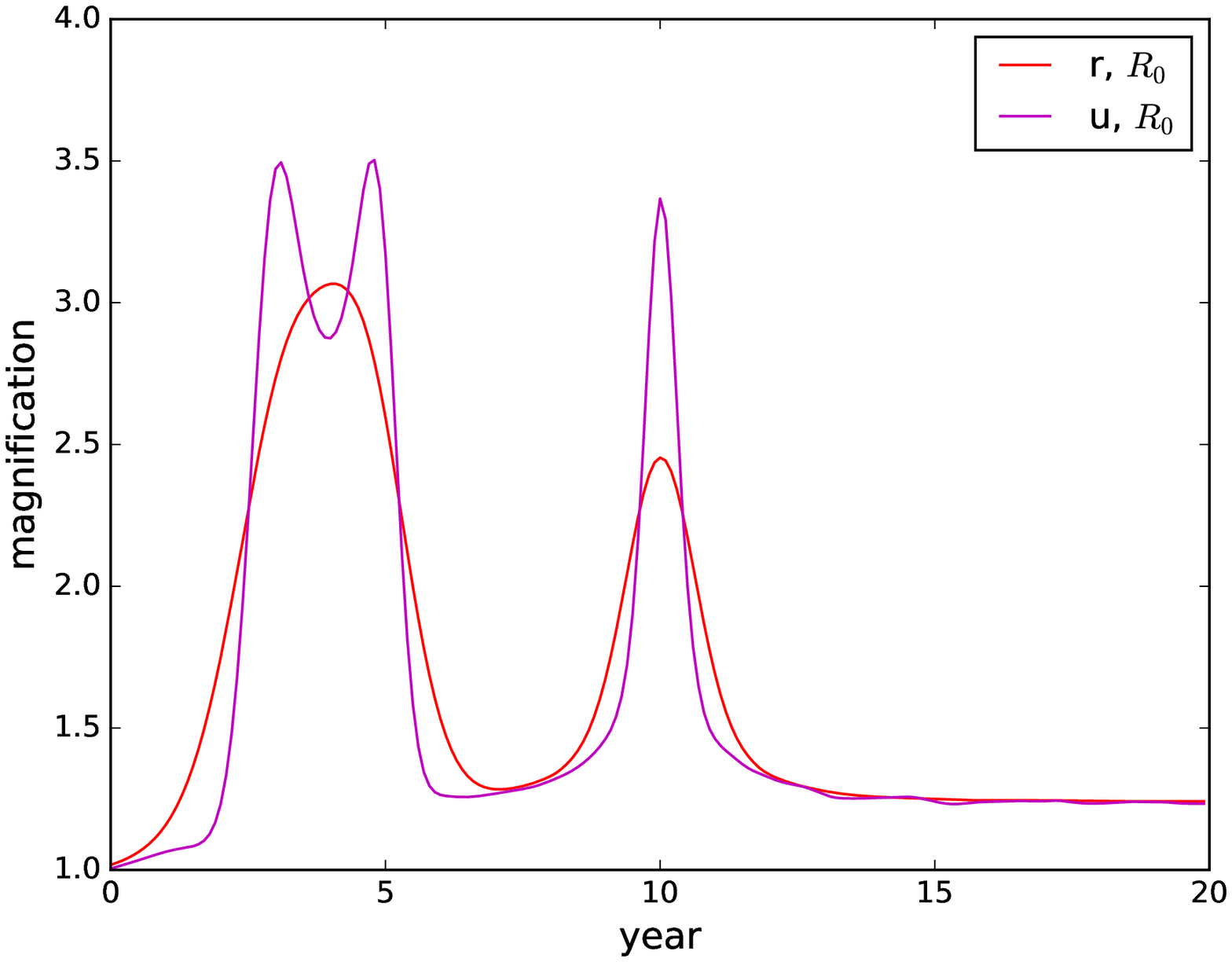}
\includegraphics[width=5cm,angle=0]{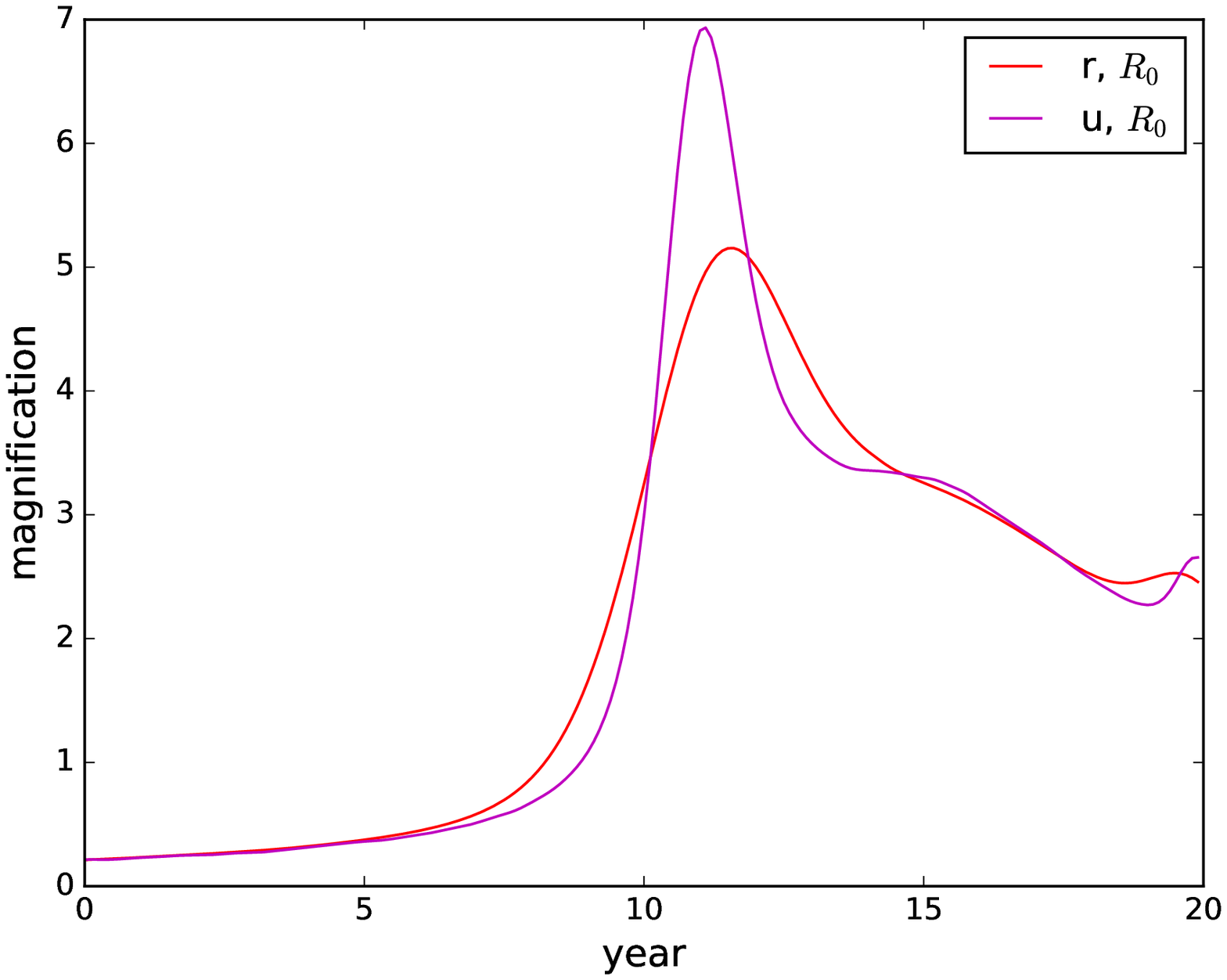}
\includegraphics[width=5cm,angle=0]{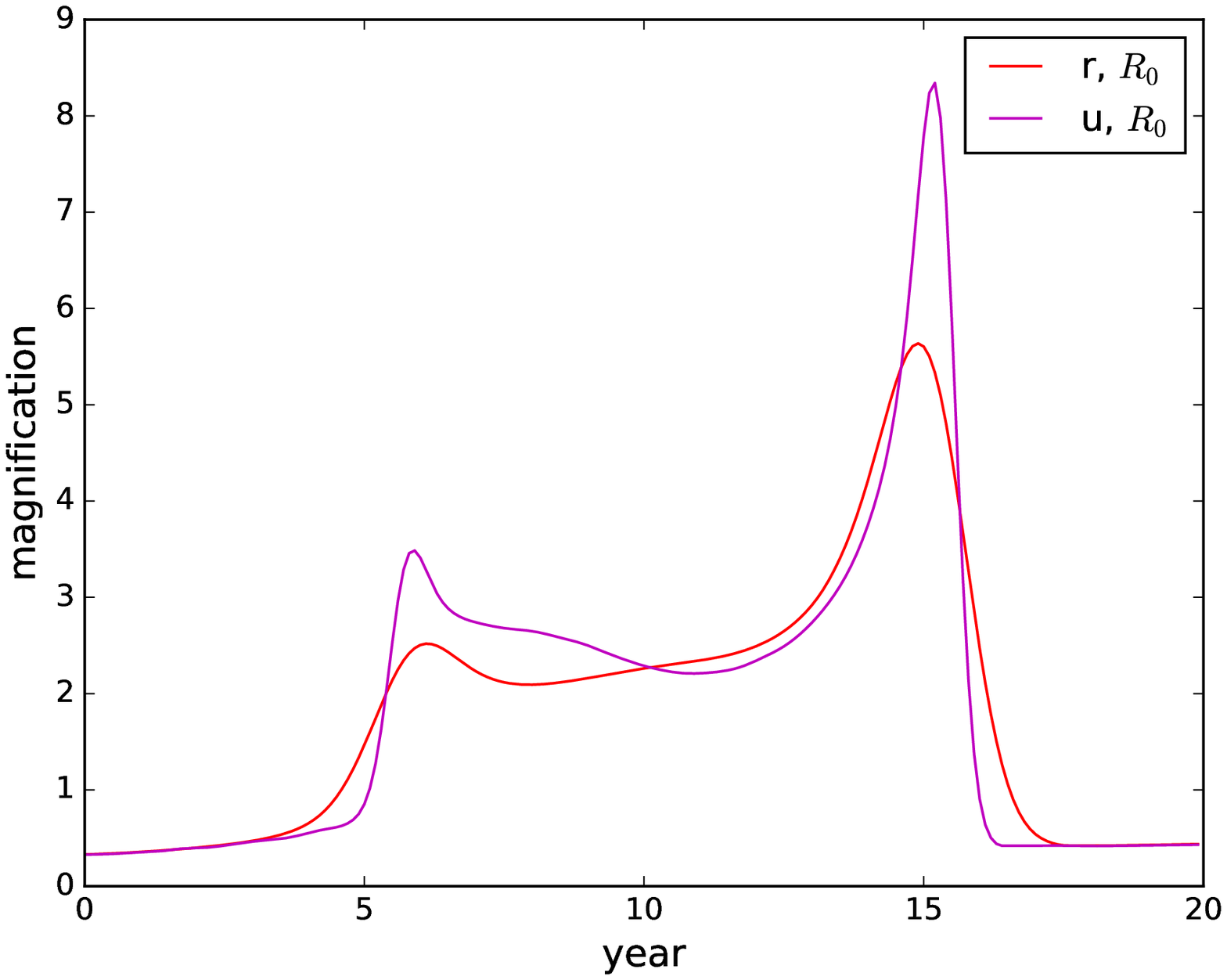}
\includegraphics[width=5.15cm,angle=0]{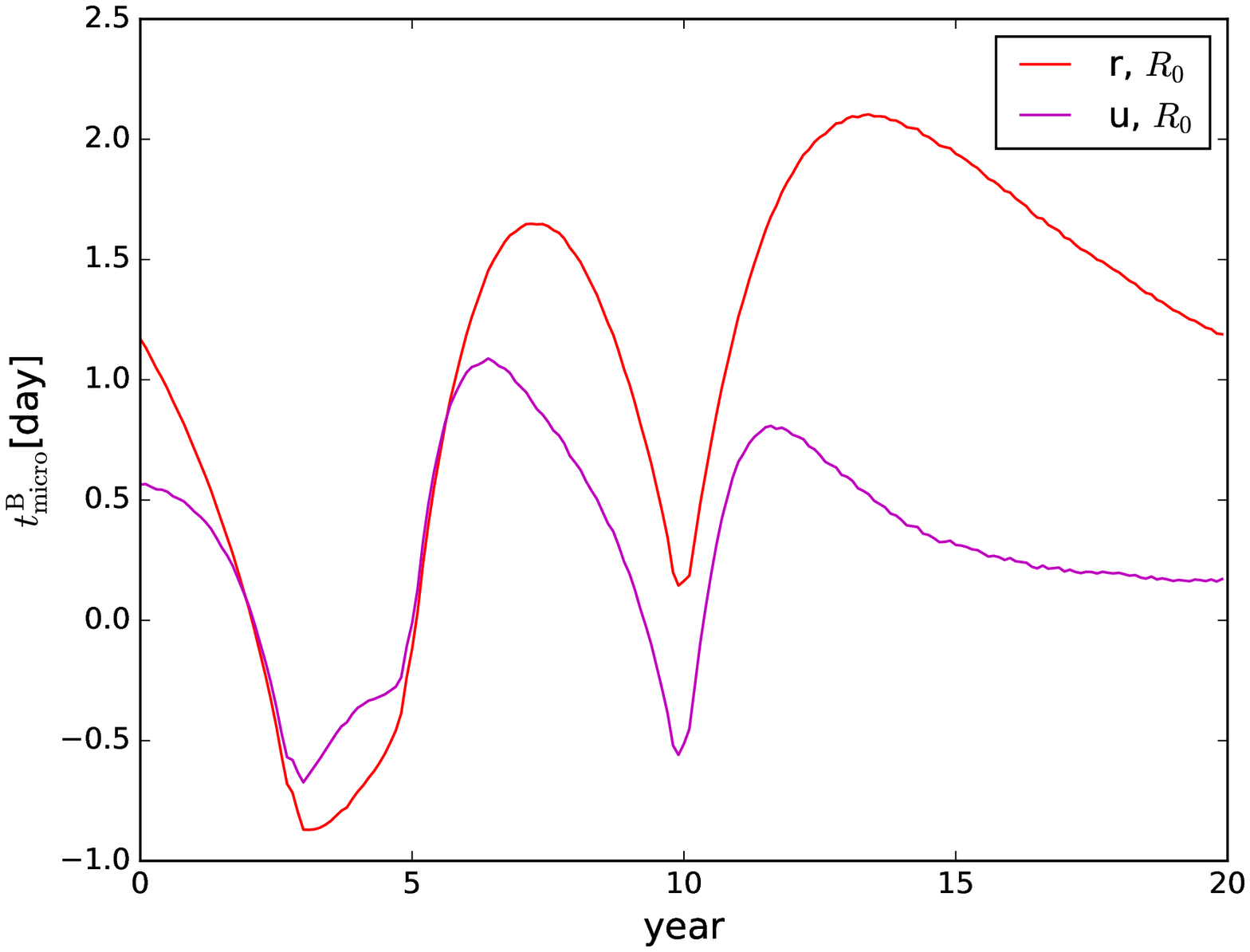}
\includegraphics[width=5cm,angle=0]{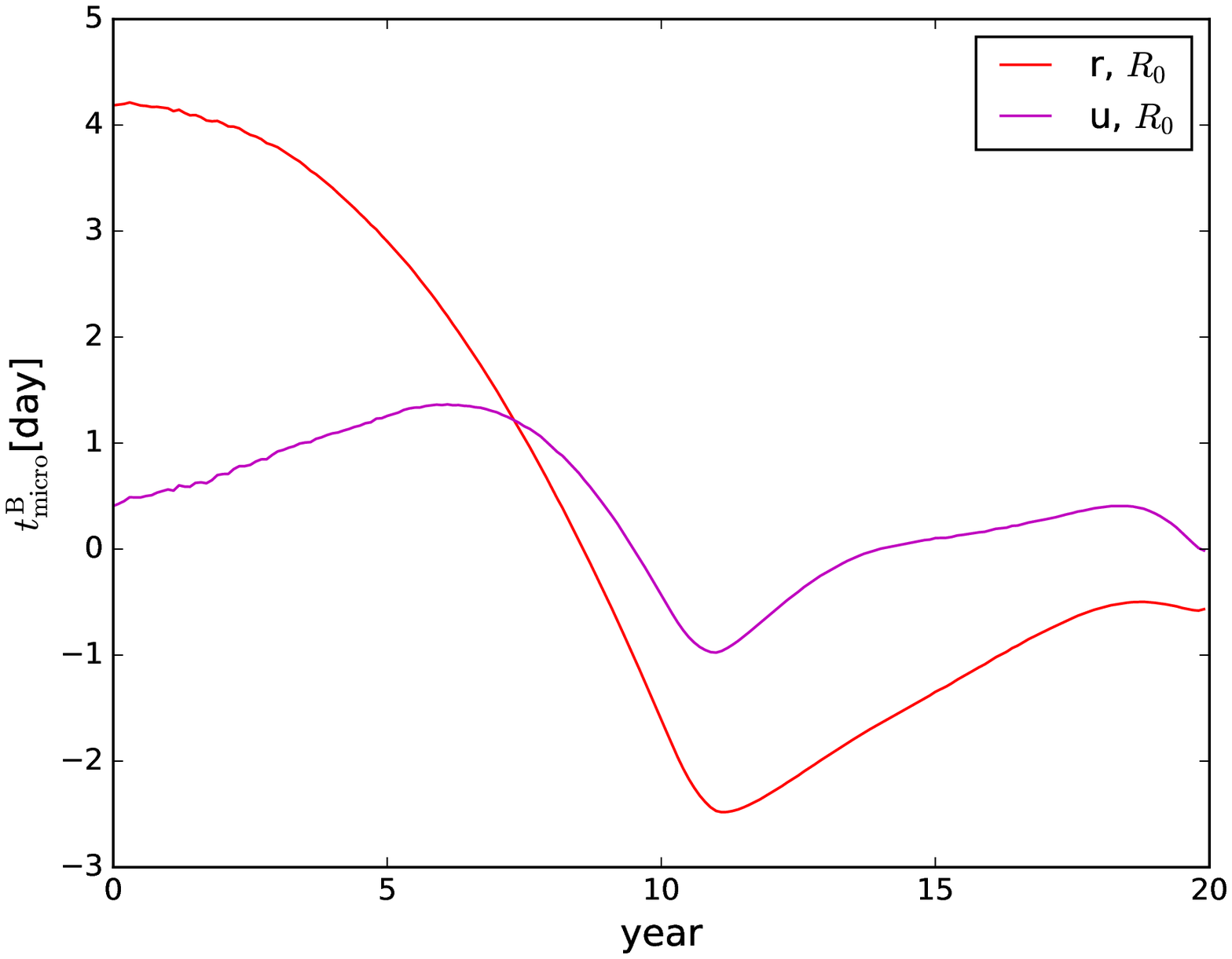}
\includegraphics[width=5cm,angle=0]{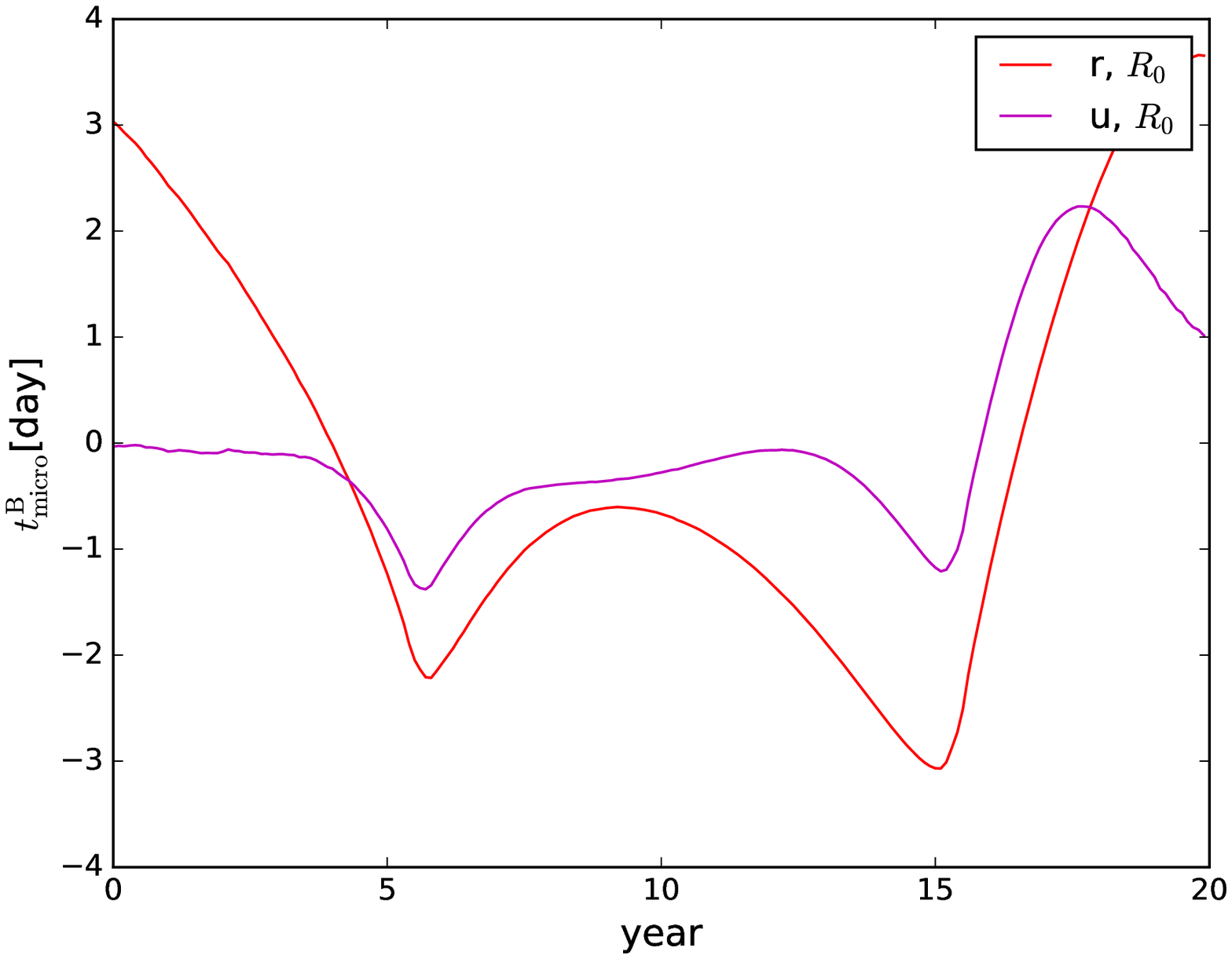}
\caption{Examples of image B with disk size $R_0$. Upper panels: microlensing magnification curves in the two bands; Lower panels: microlensing time-lag curves in the two bands.
}\label{mt}
\end{figure*}

To confirm our conjectures,
we do a simulation based on a PG 1115+080-like lensed quasar which has been used in our previous work to deal with the chromatic microlensing time-delays~\citep{Liao2020}.
The redshifts are $z_s=1.722$ and $z_l=0.311$ for the source and the lens, respectively. The newest time-delay measurements were given by~\citep{Bonvin2018}.
The two brightest images A1 and A2 are too close to split up and measured together in a single component called image A.
Time-delay between A and C is $\sim10$ days.
We take the largest time-delay $\Delta t^\mathrm{BC}_\mathrm{lc}\sim20$ days between image B and C for example in this work.
The reason is that microlensing time-delay effect impacts more on shorter time-delays. Current lensing programs like H0LiCOW~\citep{Wong2020} and TDCOSMO~\citep{Millon2020}
are inclined to use the largest time-delay in a system to infer $H_0$. This adoption is also consistent with ~\citep{Liao2020} for comparison.

The parameters related with microlensing
are convergence $\kappa$, shear $\gamma$ and star proportion $\kappa_*/\kappa$ for each image.
For image B, they are 0.502, 0.811 and 0.331, respectively while for image C, they are 0.356, 0.315 and 0.203, respectively.
These values are inferred by macro lens modelling and assuming the stellar mass-luminosity ratio~\citep{Chen2019}.
Note that there might be degeneracies, for example the mass-sheet transformation, in current lens modelling which bring uncertainties in these parameters. However,
they do not impact our analysis and conclusions since in this work we focus on the comparisons with differnt campaign lengths and
bands.
We use these parameters to generate microlensing magnification maps~\footnote{The microlensing code used in this work, MULES, is freely available at
https://github.com/gdobler/mules.} for the two images in Fig.\ref{map} (the blue ones).
The mean stellar mass is set to be $\langle M_*\rangle=0.3M_\odot$ and the Salpeter mass function is adopted with a ratio of the upper and lower masses being $M_\mathrm{upper}/M_\mathrm{lower}=100$.
Note that the presence of planet-mass lenses~\citep{Dai2018,Bhatiani2019} might potentially impact the time delay calculations, which is worth further studying.
Currently we only consider stellar lenses which dominate.
The maps have a size of $20\langle R_\mathrm{Ein}\rangle \times 20\langle R_\mathrm{Ein}\rangle$ with a pixel resolution of $4096\times4096$,
where the mean Einstein radius $\langle R_\mathrm{Ein}\rangle=3.6\times10^{16}$cm in the source plane. To avoid the impacts of specific realizations by the random seeds used to generate these maps,
we actually generate multiple such maps for each image in order to get the average distributions.

Then we calculate the microlensing time-lag maps using Eq.\ref{tmicro}. We firstly consider a standard thin-disk model with the disk size $R_0$ in Eq.\ref{R0}. In addition,
we consider a non-standard one which has twice larger disk size $2R_0$.
The disk sizes are calculated for u and r bands, respectively: $R_0(r)=1.63\times10^{15}$cm and $R_0(u)=7.24\times10^{14}$cm.
The central wavelengths are 651nm and 354nm for r band and u band in observed-frame, respectively.
Time-lag map depends on the source configuration, i.e., the inclination angle $\beta$ and the position angle PA relative to the magnification map.
However, previous studies have found these two parameters impact little on the results. To avoid disorder in the figures and tables, we only consider the case with $\beta=30^\circ$ and PA=$0^\circ$.
The time-lag maps for image B and C in r band are presented in Fig.\ref{map} (the grey ones). As one can see, the variations in the maps of image B are much larger than those in image C due to
the local environments which generate different magnification maps.
One can also notice that the time-lag maps trace the magnification maps. Therefore, the microlensing time-delays can be measured by the observed anomalies of magnification ratios
in traditional microlensing approach~\footnote{We give a brief discussion in the appendix.}.

\begin{table*}
\centering
 \begin{tabular}{lccccccc}
  \hline\hline
  band & campaign & disk size & mean &  50th percentile (median)  & 16th percentile  & 84th percentile & $\sigma$\\
  \hline
  $r$ & map & $R_0$ & 0.915 &  0.689  & -0.844  & 2.831  & 1.837\\
  $r$& 20 years & $R_0$ & 0.911 &  0.751 &  -0.095  & 1.983  & 1.039\\
  $r$& map & 2$R_0$ & 2.110 &  1.721  & -2.793  & 7.224  & 5.009\\
  $r$& 20 years & 2$R_0$ & 2.099  & 1.667 &  -1.066 &  5.534 &  3.300\\
  $u$ & map&  $R_0$ & 0.215 &  0.076 &  -0.155  & 0.638  & 0.396\\
  $u$ & 20 years&  $R_0$ & 0.212 &  0.079 &  -0.092  & 0.381  & 0.237\\
  $u$ &map &  $2R_0$ & 0.757  & 0.528  & -0.673  & 2.328 &  1.500\\
  $u$ &20 years &  $2R_0$ & 0.755  & 0.510  & -0.404  & 1.411 &  0.908\\
  \hline\hline
 \end{tabular}
 \caption{Statistics for distributions of microlensing time-delays in all cases in units of day. ``map" corresponds
 to the montionless case of the source. The dispersion $\sigma$ is defined by percentile (84th-16th)/2 for simplicity.}\label{statistics}
\end{table*}

Rather than keeping the source motionless, we consider finite lightcurves corresponding to the trace line of the motion of the source.
When calculating the time-lag curves or magnification curves, we assume Gaussian distributions for the components
of the relative velocity $v$ between the source and the star fields in the lens, with standard
deviation of 500$\mathrm{km/s}$ in each direction. The campaign is considered to be 20 years which is the typical limitation for a dedicated program.
For a 500$\mathrm{km/s}$ velocity, 20 years' motion corresponds to an angular distance of $\sim1.5\langle R_\mathrm{Ein}\rangle$.
We randomly select relative velocities, start points and directions in the maps.
Each selection corresponds to a time-lag curve and a magnification curve as well for each image.
Then we calculate the microlensing time-lags with Eq.\ref{meant}.
The distributions of the microlensing time-lags for the two images and the microlensing time-delay between them are shown in Fig.\ref{tdis}.
Statistics are summarized in Tab.\ref{statistics}. As one can see, for a 20-year campaign in r band, the time-delay dispersion can be reduced from
1.837 days and 5.009 days to 1.039 days and 3.3 days, for disk sizes $R_0$ and 2$R_0$, respectively.
The finite lightcurves reduce the time-delay dispersions caused by microlensing typically by $\sim40\%$.

Moreover, we calculate the microlensing time-delay dispersions in u band for disk sizes $R_0$ and $2R_0$, respectively.
The distributions in motionless case are shown in Fig.\ref{ru} and the statistics in all cases are summarized in
Tab.\ref{statistics}. Compared to r band, the dispersions are reduced by typically $\sim75\%$,
while the dispersions relative to the median values are similar. The extreme comparison is between
r band in motionless case and u band with 20-year monitoring. The later can reduce the dispersion by $\sim87\%$.
While bluer bands give smaller dispersions of microlensing time-delays, we want to remind the readers that the microlensing magnification curves
would have larger fluctuations and contaminate the shapes of the intrinsic variability more due to the corresponding smaller emission region with convolution.
To show the corresponding relation, we plot Fig.\ref{mt} where one can see that compared to r band,
fluctuations in u band are intenser in the magnification curve and microlensing time-lag effect is weaker (the curve is closer to zero on the whole).
One may doubt that bluer bands always reduce bias.
In the TDC, the disk sizes were simulated from $10^{14}$cm to $10^{16}$cm.
Their impacts on the accuracy of time-delay measurements were investigated and no obvious dependency was found in the results.
Therefore, blue bands like u band will not bias the results from microlensing magnification curve aspect. The reason could be that even for u band emission,
the time scales of microlensing magnification variation are still much larger than the intrinsic ones described by the damped random walk (DRW) process.

\section{Summary and discussions}

The accuracy of cosmological results becomes more and more important in the current precision cosmology era.
Systematic errors and how to control them are worth studying. For strong lens time-delay cosmography, one of the systematics comes from the time-delay measurements potentially biased by
the microlensing time-delay effect.
In summary, we have proved that increasing the monitoring time and choosing bluer band could indeed reduce the impacts by microlensing on time-delay measurements by
as large as $\sim40\%$ and $\sim75\%$, respectively. Nevertheless, the results show such an effect can not be eliminated and further studies are required to infer an unbiased $H_0$, especially
for systems like PG 1115+080 having short time-delays.

We only discuss the observation strategies for the given lens. In fact, in addition to the band and the campaign length, lens selection is important as well.
Factors like redshifts, black hole mass and local environments
of the images should be incorporated into the consideration. The angular diameter distance reaches the maximum at redshift $\sim1.6$, suggesting without considering the evolution,
quasars at this redshift have relatively the smallest observed angular disk size and thus the impact.
The $\kappa,\gamma,\kappa_*/\kappa$ of image B generate more significant effects than those of image C,
suggesting one can find lenses whose all images generate small time-lag dispersions and therefore making the time-delay measurements nearly unbiased.
In the appendix of ~\citep{Liao2015}, the standard deviation of magnification map as a function of $\kappa,\gamma,\kappa_*/\kappa$ was studied. Since the time-lag map traces magnification map,
one can easily find the best point in the parameter space that generates minimum time-lag dispersion.
Although realistic lens findings and analysis need to consider more factors, our results could be a theoretical guidance for surveys like the Rubin Observatory Legacy
Surveys of Space and Time (LSST).

\begin{figure*}
\centering
\includegraphics[width=\columnwidth,angle=0]{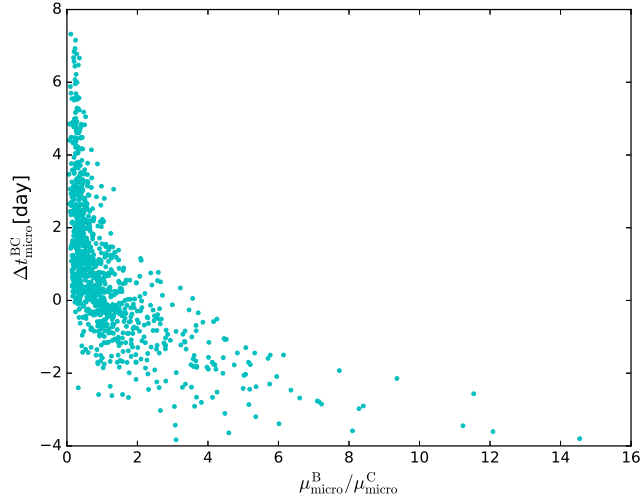}
\caption{The anti-correlation relation between microlensing magnification ratio and time-delay, for image B and C of PG 1115+080, in r band, from 1000 realization points and assuming the standard disk model.
}\label{mrt}
\end{figure*}

\section*{Acknowledgments}
I thank the referee for the helpful comments. This work was supported by the National Natural Science Foundation of China (NSFC) No.~11973034 and
the Fundamental Research Funds for the Central Universities (WUT: 2020IB022).

\section*{Appendix}

\section*{The anti-correlation between microlensing time-delay and magnification ratio}
In this appendix, we give a brief discussion on the anti-correlation between microlensing time-lag and magnification.
One can easily see the time-lags trace the magnifications in Fig.\ref{map} since the points in the
time-lag maps are determined by the corresponding local parts of magnification maps.
Note that the convolved magnification maps with finite disk size are similar as well.
Moreover, the microlensing time-delay between images should be correlated with the microlensing magnification ratio.
The latter can be measured with traditional microlensing method, i.e., to see the deviation from
the one inferred by macro lens modelling. Therefore, in addition to the time-delay ratio method~\citep{Liao2020},
microlensing time-delay can be measured with microlensing magnification ratio. To give an intuitive diagram, we take the
same lens system for example. We firstly convolve the
magnification maps with the standard disk size in r band considering all inclination and position angles of the disk.
Then we randomly select 1000 points in the maps and show the anti-correlation distribution/relation in Fig.\ref{mrt}.
This relation which gives more stringent priors on microlensing time delays from observations can be incorporated in the Bayesian framework~\citep{Chen2018} to infer an unbiased $H_0$.

\clearpage

\end{document}